\documentclass [aps, pre, amssymb, amsmath, twocolumn]{revtex4}

\usepackage{graphicx}
\usepackage{bm}

\begin{document}

\title{Transition times in the low-noise limit of stochastic dynamics}
\author{Sergey V.  \surname{Malinin}$^a$}
\author {Vladimir Y. \surname{Chernyak}$^a$}
\email{chernyak@chem.wayne.edu}

\affiliation{$^a$Department of Chemistry, Wayne
State University, 5101 Cass Avenue, Detroit, Michigan 48202, USA}
\date{\today}

\begin{abstract}
We study the transition time distribution for a particle moving between
two wells of a multidimensional potential in the low-noise limit of overdamped Langevin dynamics.
Possible transition paths are restricted to a thin tube surrounding the most probable trajectory.
We demonstrate that finding the transition time distribution reduces to a one-dimensional problem.
The resulting transition time distribution has a universal and compact form.
We suggest that transition barriers can be estimated from a single-temperature experiment if both the life
times and the transition times are measured.
\end{abstract}


\maketitle

\section{Introduction}

Dynamical processes in condensed phase often involve transitions between
metastable states. Although their timescales in different systems may differ  by
several orders of magnitude, such processes can often be described within
a common framework.

A general approach to molecular systems in condensed phase is to describe it in terms of an effective potential
energy represented by a function of the system coordinates.
The environment, represented by the degrees of freedom we are not interested in and cannot control,
provides viscosity, as well as random forces (noise) that make the observable dynamics stochastic \cite{Gardiner}.
If the case of weak noise the system spends most of time near the effective potential minima and
rarely makes transitions between potential wells.
The above scenario forms a foundation of the transition state theory for chemical reaction
rates dominated by thermoactivated processes.
Bulk experiments, where transition rates are easily observable,
motivated the development of the transition (reaction) rate theory for a variety of situations \cite{HTB}.
The transition rates can be obtained directly by observing a chemical reaction;
their ratios can be retrieved from the equilibrium concentrations.
The probability distribution functions (p.d.f.) of these life (dwell) times, i.e., the times
the system spends in potential wells between the transitions, are represented by the exponential functions,
and, therefore, are fully determined by the transition rates.

The bulk experiments, interpreted in terms of the transition rate theory, provide some information
on the system and the environment. However, the vast majority of their properties remain unexplored
due to the ensemble averaging. The latter, being automatically performed in the bulk experiments,
hides a lot of dynamical information, e.g., the details of the transition paths \cite{MO,BJS,Ritort}.
For example, within the low-temperature transition rate theory, the life times are exponentially long,
whereas the transitions themselves are treated as almost instantaneous, with their durations being irrelevant.

Development of experimental techniques with improved spatial and temporal
resolution (in particular, single-molecule experiments)
allowed the stochastic phenomena to be studied with the ensemble
averaging avoided, thus creating to a new area in statistical physics.
Theoretical effort has been mostly focused on the fluctuation theorems and related
problems (see, for example,
Refs. \onlinecite{93EvansCohenMorris,95GallavottiCohen,97Jarz,00Crooks,05Seifert,
06CCJarz,05CCJarz,LS99,98Kurchan,HSch07,TurCCP07,07CKLTur}).

In experiments with single molecules
and well-controlled systems from nano- to macro-scales
\cite{LLZTHC,TLRTJDH,SE,NGHS,AC05,LMcCD,HZRD,LMMMS,08CDS}
one can observe stochastic
trajectories that drive the system far from equilibrium, even if on longer time or length scales
the systems exhibit averaged time-independent (stationary) behavior.
Detailed information on such processes, e.g., the hydrogen-bond rearrangement,
can be also obtained via the bulk ensemble-averaged measurements, if the proper advanced spectroscopic tools,
such as 2D infrared spectroscopy \cite{Hochstrasser, CBHDCNEM, KHKSH, Tokmakoff1, Tokmakoff2, CGJT, PMF}, 
are applied.

One can study different statistical characteristics of transition trajectories
\cite{DMSSS,MaierStein,08CDS, BHB, LN, ZW, ZJZ, MT}.
In this paper we focus on the distribution of the
transition durations, which is a first passage time (FPT) problem.
For a given transition trajectory (also referred to as a switching path)
the transition time is defined is the interval between the last moment
the trajectory leaves a neighborhood of one metastable point and the first instance of time when
the trajectory enters a neighborhood of the other metastable point.
A review of work on transition-event durations as well as similar problems can be found in 
Ref. \onlinecite{ZJZ}.

Most existing theoretical models that describe multidimensional transition processes
(e.g., in Refs. \onlinecite{SOW,ZJZ,BHB,LN,BH,KT}) are one-dimensional.
Naturally, a realistic model of a complex system should include more than one degree of freedom,
and a theoretical background to justify such a reduction (identification of a 1D reaction coordinate) 
is still being developed.
In this manuscript we present
a systematic study of the transition time probability distribution function (p.d.f.) for
a multidimensional potential.
In our model the driving force in conservative, i.e., represented by the gradient of a potential function,
with the corresponding potential being smooth and not
having many energy and length scales. Since at low temperatures (or, equivalently, in the weak-noise limit)
the transitions
are rare, one can focus on a single transition between two potential wells through a saddle point 
(which may be viewed as a transition state).
Once the single transition is analyzed, the theory can be extended to complex realistic systems with
many metastable states (e.g., hydrogen-bond networks).

In this manuscript we show that in the weak-noise limit the transition time p.d.f.
has an almost universal 1D form.
In the above limit the transition time distribution is determined by a
small number of parameters, related to the potential and diffusion tensor.
This is due to the slowing down of dominating transition trajectories
when they pass the saddle point.
At this point it would be worth noting that the 1D nature of the transitions between the metastable states
originates from the effect of the transition paths being statistically restricted to the narrow tubes
that surround the most probable path
\cite{04TNK,CaroliCaroliRouletGouyet}.
It is a general phenomenon that is characteristic of low-temperature non-equilibrium and relaxation phenomena.
We have recently demonstrated \cite{CCMT09} that in the weak-noise limit the statistics of
stationary topological currents in the infinite-dimensional field theory can be found from an effective
quasi-1D Markov chain model.

The details of the transition processes can be especially important if their
durations, although being shorter than the life times (so that the metastable states can be distinguished),
are not negligibly short. This is the case of the activation barriers
being higher than $k_BT$, yet not exceeding several $k_BT$,
with $k_B$ being the Boltzmann constant. Such a situation is typical for room temperatures and hydrogen bonds,
which can be found in various systems from biomolecules to bulk and surface water.

In many cases, e.g. for biomolecules, the properties of
the thermodynamic (effective) potential and the environment are strongly
temperature-dependent due to vicinity of the melting phase transitions. This substantially
complicates the extraction of the relevant parameters, based on measuring the lifetimes at different temperatures.
As discussed in Sec. \ref{sec:discussion}, the simple universal form
of the transition time p.d.f. allows these properties to be retrieved
from single-temperature experiments, provided both the lifetimes and the transition times are measured.

The manuscript is organized as follows. In Sec. \ref{sec:paths}
we introduce our basic model of overdamped stochastic dynamics with white Gaussian noise.
In Sec. \ref{sec:trtimes}  we define the transition duration and calculate
the transition time p.d.f. Sec. \ref{sec:numerical} presents a comparison of the analytical and numerical
results for moderately weak noise.
In Appendix we derive an intuitive low-noise relation between the transition time
p.d.f. and the conditional probability density obtained from the unrestricted
Fokker-Planck equation.

\section{Transition paths}
\label{sec:paths}

We consider a system with an $m$-dimensional configuration space,
described by a set $\bm\eta = (\eta^1, \ldots, \eta^m)$ of coordinates. The Langevin equation
\begin{eqnarray}
\label{Langevin-eq}
&&
\dot\eta^i(\tau)=F^{i}({\bm\eta})+\xi^{i}(\bm\eta,\tau)
\end{eqnarray}
determines stochastic dynamics of the system in the overdamped regime. Here $F^{i}$ denotes
the deterministic (advection) component of the velocity, linearly related to the
driving force $F_j$,
\begin{eqnarray}
\label{force-velocity}
&&
F^{i}=g^{ij}F_{j}\,,
\end{eqnarray}
via the mobility tensor $g^{ij}(\bm\eta)$. The deterministic dynamics is assumed to be conservative,
i.e., $F_j=-\partial_j V$, with $V(\bm{\eta})$ being the corresponding potential.
In this manuscript we focus on a situation when the potential $V(\bm\eta)$ is bistable, i.e.,
it has two local minima (basins)
$V(\bm y_1) = V_1$ and $V(\bm y_2) = V_2$. The ways of extension of our approach to
a more general situation with multiple basins is briefly outlined in Sec.~\ref{sec:discussion}.

The mobility tensor can be viewed as a
Riemann metric in the configuration space. Due to the Einstein relation
(fluctuation-dissipation theorem), which reflects the fact that the bath is at equilibrium
with the temperature $T$, the same tensor $g^{ij}$ weighted with a factor $\kappa=k_{B}T$
characterizes the correlations for white Gaussian noise:
\begin{eqnarray}
\label{noise}
&&
\langle \xi^{i}(\bm\eta,\tau_2) \xi^{j}(\bm\eta,\tau_1) \rangle=
\kappa g^{ij}(\bm\eta) \delta (\tau_2-\tau_1)\,.
\end{eqnarray}
The dynamics represented by Eqs. (\ref{Langevin-eq}) and (\ref{noise})
allows for a path integral description with the Onsager-Machlup \cite{OM53} action
(hereafter we imply summation over the repeated indices)
\begin{eqnarray}
\label{action}
S(\bm\eta) =
(1/2)\int_{0}^{t}d\tau \,
g_{ik}\left(\dot{\eta}^{i}-F^{i}({\bm \eta})\right)
\left(\dot{\eta}^{k}-F^{k}({\bm \eta})\right),
\end{eqnarray}
where $g_{ij}$ is defined by $g_{ij}g^{jk}=\delta_{i}^{k}$.
The transition probability $K(\bm{x}'',\bm{x}';t)$, i.e. the probability for the system
to move from point $\bm{x}'$ in the configuration space to point $\bm{x}''$ over time $t$ is given by a path integral
\begin{eqnarray}
\label{path-int-transition}
K(\bm{x}'',\bm{x}';t)=\int{\cal D}{\bm\eta}e^{-S({\bm\eta})/\kappa},
\end{eqnarray}
where integration goes over all trajectories with ${\bm\eta}(0)={\bm x}'$ and ${\bm\eta}(t)={\bm x}''$.
The action in the path-integral representation [Eq.~(\ref{path-int-transition})] is understood as a proper
discrete-time action, whose continuous limit is given by Eq.~(\ref{action}) \cite{HuntRoss}.

In the weak-noise $\kappa\to 0$ limit the expression for $K(\bm{x}'',\bm{x}';t)$
[Eq.~(\ref{path-int-transition})] suggests that a transition is dominated by the trajectory
that minimizes the action $S({\bm\eta})$. It is well-known that integration in Eq.~(\ref{path-int-transition})
goes over continuous, rather than smooth trajectories \cite{Feynman}, which can be actually considered
as a reason why the action in Eq.~(\ref{action}), defined for smooth trajectories,
requires regularization (i.e., a choice of a proper discrete-time form).
However, in the limit of a fine discretization (small time step) the dominant trajectory becomes smooth,
and, therefore, can be found from the Onsager-Machlup action (\ref{action}) by solving the corresponding
Euler-Lagrange equation.

There are two special types of trajectories that provide local minima of the Onsager-Machlup action.
Its absolute minima ($S=0$) correspond to the trajectories that satisfy $\dot{\eta}^i=F^i({\bm \eta}(t))$.
However, for such downhill trajectories the value of the potential decreases monotonically with time,
whereas a transition between two basins involves the potential increase at the first stage of the process.
In the conservative $F_{j}=-\partial_{j}V$ case under consideration an uphill trajectory that satisfies
$\dot{\eta}^i=-F^i({\bm \eta}(t))$
also satisfies the Euler-Lagrange equation. Note that an uphill trajectory is nothing else
that a time-reversed counterpart of a downhill trajectory.

The scenario of a transition between two basins has been thoroughly studied in the context of
the reaction rate theory \cite{Gardiner}. The system follows the uphill trajectory
to reach the transition state (a critical point of the potential with a single unstable mode) and
then reaches the other basin via the downhill trajectory. The uphill and downhill parts
coming to rest at the saddle point constitute a special trajectory
(hereafter referred to as the reference trajectory) that minimizes the Onsager-Machlup action.
The downhill part does not need any noise,
whereas for the uphill part certain concerted noise is necessary to overcome the deterministic force.
The two parts of the reference trajectory satisfy the equations
\begin{eqnarray}
\label{downhill-uphill}
&&
\dot{\eta}^i=\pm F^i({\bm \eta}(t))\,.
\end{eqnarray}

The reference trajectory is a special case of the most probable escape path (MPEP)
or the most probable switching path (MPSP) that corresponds to infinite transition time.
We assume that the reference trajectory connecting two potential minima is unique.
When the transition time is finite, the most probable trajectory deviates from its reference counterpart.
In particular, in the limit of very short times it becomes a geodesic line, and is independent of the potential.

Within the path integral approach we need to consider all possible transition trajectories.
In the weak-noise case the situation is much simpler,
since the deviations from the reference trajectory are suppressed.
This can be used, for example, for generalizing 1D results \cite{Weiss,CaroliCaroliRoulet_pathint}
to the multidimensional case, by linearizing the theory around the reference trajectory.
Thus, we have a linear but still multidimensional
time-dependent problem. Therefore, the deviations of the MPSP from the reference trajectory due to finite
transition times, as well as Gaussian fluctuations around the MPSP may not be treated explicitly within
the analytical framework.

To avoid the aforementioned difficulties we invoke a ``Hamiltonian'', rather than ``Lagrangian'' approach:
Instead of following individual stochastic trajectories, which is the natural basis
for the path-integral approach, we reformulate the problem in terms of the probability density evolution.
The probability $dP$ to find a system in a certain volume of the configuration space is obviously an
integration measure, and does not depend on the particular choice of coordinates we use to parameterize
the configuration space. Therefore, the probability density is a scalar if defined as $\rho = dP/d\mu$
with respect to the invariant volume element $d\mu=\sqrt{g}dx^1\ldots dx^m$, where $g = \det g_{ik}$.
With this invariant definition of the probability density, the Fokker-Planck operator,
which governs its evolution, is naturally invariant with respect to coordinate transformations.

The Langevin equation (\ref{Langevin-eq}) corresponds to the following Fokker-Planck (FP) equation:
\begin{eqnarray}
\label{FP-eq-covariant}
&&
\partial_t \rho = {\cal L}\rho\,,
\end{eqnarray}
where the FP operator has a covariant form \cite{Graham}
\begin{align}
\label{FP-operator}
\nonumber
&
{\cal L}\rho = (\kappa/2)\nabla^2\rho - {\rm div\,}(\bm F\rho) =
\\
&
(\kappa/2)
(1/\sqrt{g})\partial_i\left(\sqrt{g}g^{ik} \partial_k\rho\right)
+ (1/\sqrt{g})\partial_i \left( \sqrt{g}  F^i\rho \right)
\,.
\end{align}

\section{Transition time distribution in the weak-noise limit}
\label{sec:trtimes}

A problem of the transition duration and its distribution can be formulated for any noise strength.
For strong noise the framework of overdamped dynamics with white Gaussian noise,
introduced above, may be inadequate, and a model with more parameters might be necessary.
Therefore, we restrict our treatment to the low-noise limit.

In this limit a conservative system has the following properties
\cite{LandauerSwanson,CaroliCaroliRoulet,CaroliCaroliRouletGouyet}.
First, the quasiequilibrium density distributions are strongly localized in the
potential wells, and the inter-well relaxation requires exponentially long times.
The shapes of these distributions are determined by a small number of parameters that characterize
the potential in the wells. The relaxation rates, in addition, depend on the mobility and the potential
at the transition states represented by the saddle points.
Second, the trajectories that lead from one potential well to another are also determined
by a few parameters of the deterministic drift, as we will see below.
This universality that takes place in the low-noise limit makes the model especially simple and attractive,
even though it can be imprecise when applied to stronger noise.

The analysis of the Onsager-Machlup action or the FP operator shows that a characteristic
transverse deviation of the trajectory from the most probable trajectory is $\propto\sqrt{\kappa}$.
For weak noise, most observed transition trajectories are close
to the reference trajectory, since they typically correspond to long transition times,
as follows from the results of Subsection \ref{subsec:pdf-calculation}.
Thus, all probable transition trajectories can be
considered to be enclosed in a thin tube of the width $\propto\sqrt{\kappa}$. The tube
extends from one potential well to the other and is represented by a tubular neighborhood
of the reference trajectory. The concept of a tube,
as a region outside of the potential minima where the transition density is localized,
appeared, for example, in Refs.
\onlinecite{LandauerSwanson,CaroliCaroliRouletGouyet,DMRH,MaierStein2,DLMcCS,04TNK,08CDS,CCMT09}.

\subsection{Coordinates in the transition tube}
\label{subsec:tubecoordinates}

Once the reference trajectory
[the MPSP corresponding to the infinite transition time defined in Eqs. (\ref{downhill-uphill})]
in the $m$-dimensional configuration space is known, we can introduce a coordinate system in the tube
that allows all possible transition trajectories to be described  in a convenient and natural way.
The coordinates are represented by the distance $x$ along the reference trajectory and a set
of $(m-1)$ mutually orthogonal transverse coordinates $\bm\zeta$.
We set $x=0$ at the saddle point and denote the longitudinal coordinates of the first and second minima
by $y_1<0$ and $y_2>0$, respectively.
The transverse coordinates can be selected in the following way.
For a point on the reference trajectory located at $x$ and a tangent vector $\bm{\zeta}$
(i.e., an infinitesimal shift of a configuration) at this point, which is normal to the reference trajectory,
we consider a length $|\bm{\zeta}|$ segment of a geodesic line (recall that the mobility tensor
defines a Riemann metric in the configuration space) that starts at the point $x$ on the reference trajectory
in the direction of ${\bm\zeta}$. The end point of the segment is denoted by $(x,\bm{\zeta})$.
For small enough $|\bm{\zeta}|$ (actually small compared to the trajectory curvature radius)
the sets $(x,\bm{\zeta})$ unambiguously identify the configurations in the tubular neighborhood
of the reference trajectory. At this point a global coordinate system in the tube can be built by choosing
the basis sets $\bm{e}_{1}(x),\ldots,\bm{e}_{m-1}(x)$ in the orthogonal spaces to the reference trajectory
with a smooth dependence on $x$. Naturally a coordinate set $(x,\zeta^{1},\ldots,\zeta^{m-1})$
represents the point $(x,\bm{\zeta})$ with $\bm{\zeta}=\sum_{\alpha=1}^{m-1}\zeta^{\alpha}\bm{e}_{\alpha}(x)$.

Since we are dealing with a weak-noise (low-temperature) and long-time situation, so that the
deviations from the reference trajectory are small, we can expand all relevant quantities
in the deviations $\bm\zeta$ from the reference trajectory.
Thus, the potential in the vicinity of the tube is given by
\begin{align}
\label{eq:potential}
&
V(\bm\eta)=V_0(x)+(1/2)W_{\alpha\beta}(x)\zeta^\alpha\zeta^\beta\,.
\end{align}
The terms linear in $\bm \zeta$ do not appear in the expansion, since the potential
gradient is directed along the reference trajectory. In the low-noise limit
the expansion to second order in $\bm\zeta$ is sufficient, provided $W$ is non-degenerate,
i.e., $W$ is positively defined along the transition trajectory.
Therefore,
we will need the potential only the form given by Eq. (\ref{eq:potential}), and in particular use the
longitudinal force on the reference trajectory
\begin{align}
\label{F0}
&
F_0(x)=-\partial_x V_0(x)\,,
\end{align}
which has zeros in the stationary points $y_1$, $0$, and $y_2$.

\subsection{Definition of transitions}
\label{subsec:trans-def}

\begin{figure}[tp]
  \begin{center}
      \includegraphics[width=0.4\textwidth]{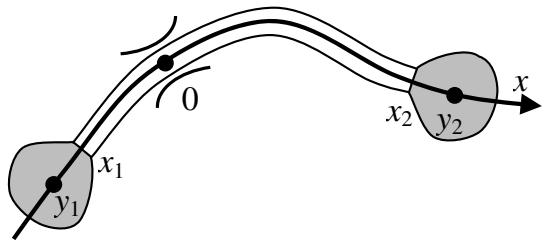}
  \end{center}
  \caption{Most transition trajectories between neighborhoods of the stable points are within
  a transition tube.}
  \label{fig:tube}
\end{figure}

We are focused on a situation when the particle makes a transition from one potential
well to another via a single ``channel" that passes through the potential saddle point.
We assume the temperature (noise) to be small compared to the height of the potential hill.
Thus, a particle located in a potential well has enough time to reach an almost equilibrium distribution
before a rare strong-noise fluctuation, represented by a series of concerted random force
kicks, pushes it over the potential saddle point to the other potential well.
A transition is naturally defined as an event when a particle leaves a neighborhood $U_1$ of
the initial basin, located at $x=y_1$, and enters a neighborhood $U_2$ of the other minimum
located at $x=y_2$. Thus, the transition time is the first passage time \cite{Risken} from the initial
boundary $\partial U_1$ to the final boundary $\partial U_2$ without revisiting the initial boundary.
It is convenient and natural to choose the regions $U_1$ and $U_2$ in such a way that the probability
to find the particle outside of them is small. On the other hand, since we study the transitions
between metastable states, the regions $U_1$ and $U_2$ should be small compared to the
typical length scale of the problem. These two conditions are compatible, provided the noise strength
$\kappa$ is small, compared to the barrier height. The subsequent treatment of the problem can be
simplified without losing the essential properties of its solution, if we assume the faces of the tube at
$x=x_1$ and $x=x_2$ to be parts of the boundaries $\partial U_1$ and $\partial U_2$,
respectively (see Fig. \ref{fig:tube}).
Therefore, the explicit conditions for $x_1$ and $x_2$ read:
$\sqrt{-\kappa/\partial_x F_0(y_1)} \ll (x_1-y_1)\ll 1$ and
$\sqrt{-\kappa/\partial_x F_0(y_2)} \ll (y_2-x_2)\ll 1$,
where the coordinate is rescaled, so that the typical length scale of the potential can be denoted by $1$.

The transition time p.d.f. can be represented as a path-integral form,
\begin{align}
\label{path-int}
F(t) = \int\limits_{\bm\eta(0)\in\partial U_1}^{\bm\eta(t)\in\partial U_2}
{\cal D}\bm\eta e^{-S(\bm\eta)}\,,
\end{align}
with the integral measure including only those trajectories $\bm\eta(\tau)$ that
satisfy the initial and final conditions, specified in the integral and
do not visit $U_1$ or $U_2$ at $\tau\in(0,t)$.
As already mentioned in Sec.~\ref{sec:paths} the path integral can be defined once
a proper time discretization is introduced. If time discretization involves $n$ intermediate
time points between $0$ and $t$, the regularized integration is represented by $n$ integrals
over $\bm\eta\in M\backslash(U_1\cup U_2)$, an integral over $\partial U_1$ with a proper distribution
of the starting points, and an integral over $\partial U_2$. The presented above discretized form of the
path integral has a direct interpretation in terms of the experimental observables.
Namely, in an experimental set-up one can monitor the particle's position at certain discrete times and
register the crossing of the boundary if the particle appears on its other side after a time step.

We focus on the transition from one basin to the other; in the low-noise case they occur by passing
the relevant saddle point. Since the noise is weak compared to the deterministic force on the tube faces
at $x_1$ and $x_2$, the probability of the excursions into $U_1$ or $U_2$ is low for a trajectory
that starts at $\partial U_1$ and ends at $\partial U_2$ after a relatively long time.
Therefore, we can approximate the expression (\ref{path-int}), which is exact but difficult
to deal with, by the matrix element
\begin{align}
\label{F-matrix-element}
F(t) \propto \langle \rho' | e^{{\cal L}t}|\varrho \rangle
\end{align}
with the Fokker-Planck operator ${\cal L}$ given by Eq. (\ref{FP-operator}).
The proportionality sign means that the distribution should be normalized,
since not all trajectories that start at $\partial U_1$ will reach
$\partial U_2$ (actually, only an exponentially small fraction of them will do). However, as
explained above, the trajectories that do reach $\partial U_2$ pass through the tube of the small width
$\sim\sqrt{\kappa}$ and mainly stay within the interval $(x_1,x_2)$.
In Eq. (\ref{F-matrix-element}) $|\varrho\rangle$ represents the distribution of the starting points
at $\partial U_1$, and $\langle \rho' |$ stands for the integration over the final surface.
The initial distribution has an equilibrium form and at $x=x_1$ is contained within the face of
the tube:
\begin{align}
\label{init-cond}
&
\varrho(\bm\eta) = Ze^{-\kappa^{-1}W(y_1)(\bm\zeta\otimes\bm\zeta)}\delta(x - x_1),
\end{align}
where we made use of the fact that $x_1$ is close to the minimum at $x=y_1$.

In Appendix \ref{appendix:reduction} we derive the approximation represented by 
Eq. (\ref{F-matrix-element}) in the 1D case, which can be generalized to a
multidimensional situation.

\subsection{Eigenstates of ${\cal L}$}
\label{subsec:eigenstates}

The matrix element in Eq. (\ref{F-matrix-element}) can be found as a spectral decomposition
in terms of the eigenstates of the FP operator, which are the solutions of the eigenvalue problem
${\cal L}\rho = \lambda \rho$. The calculation below can be viewed as a simplified
version of the multidimensional calculation reported in Ref. \onlinecite{CaroliCaroliRouletGouyet},
adapted to the initial distribution near the potential minimum.
We skip some finer details (irrelevant in the low-noise limit) of the derivation replacing them with
physical intuition; these details can be found in Refs. \onlinecite{CaroliCaroliRoulet,CaroliCaroliRouletGouyet}.

We seek for the eigenstates in the form
\begin{align}
\label{rho0-ansatz}
\rho(\bm\eta) = \rho_0(x) \exp(-\kappa^{-1}\sigma(x)(\bm\zeta\otimes\bm\zeta))
\end{align}
with an $x$-dependent matrix $\sigma$. Together with the noise strength $\kappa$, the matrix $\sigma$ determines
the density  distribution in the transverse direction. The resulting eigenvalue problem for $\rho_0(x)$
is represented by
\begin{align}
\label{eq:rho0}
&
(\kappa/2)\partial^2_x\rho_0 - \partial_x(\rho_0 F_0)
-\rho_0 \mathrm{Tr} (\sigma-W)  = \lambda \rho_0\,.
\end{align}

There are two kinds of regions along the tube:
a domain where the diffusion terms are small compared to the deterministic counterparts,
hereafter referred to as the WKB domain (using the analogy \cite{Risken} between the Fokker-Planck
and Schr\"odinger operators in the conservative case under study),
and the fluctuation regions located near the potential stationary points.
In the weak-noise $\kappa\to 0$ limit, the size of the fluctuation regions is proportional
to $\sqrt{\kappa}$. Therefore, the WKB domain overlaps with the region $|x|\ll 1$ near
the saddle point where the force is linear in the deviations (or the potential is harmonic,
so that the fluctuation regions can be naturally referred to as the harmonic regions).
In addition to Eq. (\ref{eq:rho0}), the eigenvalue problem ${\cal L}\rho = \lambda \rho$
leads to an equation for $\sigma(x)$, which has a simple form in the WKB domain:
\begin{align}
\label{eq:sigmaWKB}
&
2\sigma^2-\sigma W - W \sigma + F_0\nabla_x\sigma = 0
\,.
\end{align}
Here $\nabla_x$ denotes a long (covariant) derivative with respect to the connection
that takes into account the fact that the tensor $\sigma(x)$ is defined in different spaces at different $x$.
Note that the configuration-dependent mobility tensor (metric) implicitly enters
Eqs. (\ref{F0}), (\ref{eq:rho0}), and (\ref{eq:sigmaWKB}) through
the matrix operations and the longitudinal coordinate $x$ chosen as a length.
In the derivation and results below we use the potential in terms of $x$.

We will avoid explicit analysis of $\sigma$ outside the WKB domain and use
its following properties without a rigorous proof (see, for example,  \cite{CCMT09} for some details).
Close to the critical points of the potential we neglect deviations of $\sigma$ from
$W$ at the corresponding fixed points. In particular, in the saddle point $\sigma(0) = W(0)$, and
on the faces of the tube $\sigma(x_1) \approx W(y_1)$ and $\sigma(x_2) \approx W(y_2)$.
Thus, the initial distribution (\ref{init-cond}) corresponds to the ground state in the transverse direction
represented by a Gaussian. This excludes the transverse excited states from the expansion of $e^{{\cal L}t}\varrho$
over the eigenstates of ${\cal L}$. In addition, transverse excited states do not make any
contribution because the corresponding contributions vanish upon the integration performed at $\partial U_2$.

At this point we would like to emphasize that although, as shown above, the higher transverse modes
do not contribute to the transition time p.d.f., the full transition probability
$K({\bm x}'',{\bm x}';t)$ involves the complete expansion, and the higher transverse modes are not
suppressed by any small parameter. This is the ``Hamiltonian'' signature of the fact, briefly discussed
at the end of Sec.~ \ref{sec:paths} in terms of the ``Lagrangian'' path-integral language,
that although the transition amplitude is dominated by the most probable trajectory,
it may not be found explicitly in analytical terms due to the complexity of the relevant multidimensional
linear problem. This demonstrates the advantages of the ``Hamiltonian'' approach based on the Fokker-Planck
equation. This also demonstrates the non-trivial nature of the 1D-reduction of the transition-time
distribution problem, as resulting from exact cancelations related to higher-order transverse modes.

Next, we notice that the choice of $x_1$ and $x_2$
described in the subsection (\ref{subsec:trans-def}) leads to the same first-order equation for
$\sigma$ for each longitudinal mode (that corresponds to the ground-state transverse mode),
valid almost everywhere on the interval $(x_1,x_2)$, as we can see from Eq. (\ref{eq:sigmaWKB}).
Therefore, $\sigma(x)$ is the same for all eigenfunctions; it is represented by the solution
of the same equation (\ref{eq:sigmaWKB}) with the same boundary condition $\sigma(x_1) = W(y_1)$.
These arguments justify our ansatz represented by Eq.~(\ref{rho0-ansatz}).

The differential operator in Eq. (\ref{eq:rho0}) is not self-adjoint (Hermitian) but can be made
self-adjoint by applying the transformation
\begin{align}
\label{rho0psi0}
&
\rho_0(x)=\psi_0(x)\exp(-\kappa^{-1}V_0(x))\,.
\end{align}
The transformed eigenfunctions $\psi_0(x)$ satisfy the equation
\begin{align}
\label{eq:psi0}
\nonumber
&
(\kappa/2)\partial^2_x\psi_0 -
(\kappa^{-1}/2)\psi_0\left(\kappa\partial_x F_0 +F_0^2 +\mathrm{Tr} (\sigma-W)\right)
\\
&
= \lambda \psi_0
\end{align}
that can be viewed as the Schr\"odinger equation with an effective potential with deep minima near
the fixed points (stable, as well as unstable) of $V(\bm\eta)$.
The eigenfunctions responsible for transition dynamics are concentrated at the minimum of
the effective potential that corresponds the saddle point of $V(\bm\eta)$.

To calculate the transition time p.d.f. we expand the matrix element
(\ref{F-matrix-element}) of the FP evolution operator in the eigenmodes of $\cal L$ as
\begin{align}
\label{eq:rhopsi0psi0}
\nonumber
&
\langle \rho' | e^{{\cal L}t}|\varrho \rangle= Ze^{\kappa^{-1}(V_0(x_1)-V_0(x_2))}
\\
&
\times \sqrt{\frac{(\pi\kappa)^{m-1}}{\det W(x_2)}}
\sum_\lambda \psi_{0\lambda}(x_1)\psi_{0\lambda}(x_2)
e^{\lambda t}
\,.
\end{align}
Note that only the eigenstates localized in the transition region provide non-negligible contributions
to the expansion, since only these eigenstates are ``produced'' by transition trajectories that do not visit
the neighborhoods of the potential minima.

The next step requires solution of the eigenvalue problem, given
by Eqs. (\ref{eq:psi0}) and (\ref{eq:sigmaWKB}). Since the relevant eigenstates are localized at the
saddle point, it is sufficient to consider the harmonic region that contains the saddle point $x=0$,
as well as the two surrounding (and overlapping with it) WKB sub-domains.

In principle, in each WKB sub-domain the eigenfunction $\psi_{0\lambda}$ contains two ``waves''
that correspond to two linearly independent solutions of the second-order differential equation.
One ``wave'' decays as $|x|$ increases, whereas the other one grows.
For the eigenstates under study (localized near the saddle point) the growing ``wave'' can be neglected immediately.
This implies that functions $\rho_{0\lambda} = \psi_{0\lambda} e^{-V_0(x)/\kappa}$ change slowly with $x$.
Therefore, the shortest way to obtain the solution
is to neglect the second derivative in Eq. (\ref{eq:rho0}) and use Eq. (\ref{eq:sigmaWKB})
to exclude $\sigma(x)$. The resulting first-order equation
\begin{align}
\label{WKB-wave}
&
\partial_x( F_0\rho_{0\lambda}/\sqrt{\det\sigma})=
-\lambda\rho_{0\lambda}/\sqrt{\det\sigma}
\end{align}
can be easily integrated, which involves one constant:
\begin{align}
\label{WKB-wave-solution}
&
\rho_{0\lambda}(x)=D_\lambda \frac{\sqrt{\det\sigma(x)}}{F_0(x)}
\exp\left( -\lambda\int^x \frac{dz}{F_0(z)}\right)\,.
\end{align}

In the harmonic region $|x|\ll 1$, Eq. (\ref{eq:psi0}) can be represented as
\begin{align}
\label{eq:psi0-harm}
&
(\kappa/2)\partial^2_x\psi_0 -\psi_0 k^2 x^2/(2\kappa)
= (\lambda+k/2) \psi_0\,,
\end{align}
where we approximate $\sigma(x) \approx W(x) = W(0)$ and denote
\begin{align}
\label{k}
&
k = \partial_x F_0(0)
\,.
\end{align}
Since in the WKB domain there is only one ``wave'' that decays as $|x|$ increases, the eigenfunctions
in the harmonic region have the form
\begin{align}
\label{harmonic-wave}
&
\psi_{0}= \left(\frac{2^{2-2n}k}{\pi\kappa(n-1)!^2}\right)^{1/4}
\!\!
H_{n-1}\left(x\sqrt{k/\kappa} \right)e^{-k x^2/(2\kappa)},
\end{align}
whereas the eigenvalues are given by $\lambda = -nk$, with integers $n\ge 1$ and
Hermite polynomials $H_n$. The normalization factor in Eq. (\ref{harmonic-wave}) is
determined by the form of the solution in the harmonic region where it is localized.

Now we can match the solutions in the two WKB sub-domains with the solution in the harmonic region.
We will only consider the second WKB sub-domain from $x\sim\sqrt{\kappa}$ to $x=x_2$; the treatment
of the first WKB sub-domain is similar.
If we choose $x_2$ as the lower limit of the integral in Eq. (\ref{WKB-wave-solution}),
the WKB solution with $\lambda=-nk$  can be represented as
\begin{align}
\label{WKB-wave-result}
&
\rho_{0\lambda}(x)=D_\lambda k^{-1} \sqrt{\det W(0)} x^{n-1} (K/x_2)^n\,,
\\
\label{WKB-wave-result-1}
&
K = \exp\left(\int\limits_{0}^{x_2}dx \left(1/x - k/F_0(x)\right)\right)
\,
\end{align}
at $x\ll 1$. This solution should be matched with the solution $\rho_{0\lambda}(x)$, given by
Eqs. (\ref{rho0psi0}) and (\ref{harmonic-wave}), at $x\gg \sqrt{\kappa/k}$ in the harmonic region.
To achieve that, we use the asymptotic form $H_n(z) = (2z)^n$ of the Hermite polynomial, for $z\gg 1$.

Eventually we obtain the following expression for the eigenfunction values at the right boundary:
\begin{align}
\label{boundary-value}
\nonumber
&
\rho_{0\lambda}(x_2)=e^{-V_0(0)/\kappa}\sqrt{\frac{\det W(x_2)}{\det W(0)} }
\left( \frac{k}{\pi\kappa} \right)^{1/4}
\times
\\
&
\left( \frac{2k}{\kappa} \right)^{(n-1)/2}
\frac{k(x_2/K)^n }{\sqrt{(n-1)!}F_0(x_2)}\,.
\end{align}
The expressions for $\rho_{0\lambda}(x_1)$ are derived in a similar way.

\subsection{Calculation of the transition time p.d.f.}
\label{subsec:pdf-calculation}

We are now in a position to complete the calculation of the
transition time p.d.f. We first note that its long-time asymptotic form can be easily found
once we know the lowest eigenvalue of $-\cal L$. The corresponding
eigenfunction is localized in the vicinity of the saddle point, where the potential
is harmonic. Therefore, the asymptotic form $F(t)\propto e^{-kt}$ is fully determined by the curvature $k$
of the potential $V_0$ (the derivative of the longitudinal deterministic drift) at the saddle point.
However, since the expansion over the eigenstates of $\cal L$ is singular in $\kappa$ (as shown below),
this first term in the asymptotic expansion may not be used to determine $F(t)$ in a broader region,
i.e., around the most probable transition time.

We complete the derivation by collecting all terms in the expansion (\ref{eq:rhopsi0psi0}).
The time dependence of the transition time p.d.f. is given by
\begin{align}
\label{F-series-result}
F(t)  \propto \sum\limits_{n=0}^{\infty} \frac{1}{n!}\left( -2k G/\kappa \right)^n
e^{-k(n+1)t}
\,,
\end{align}
where $G$ depends on $x_1$ and $x_2$ as
\begin{align}
\label{G}
G = k|x_1|x_2 \exp\left(\int\limits_{x_1}^{x_2}dx\left(k/|F_0(x)| - 1/|x|\right)\right)
\,.
\end{align}
We further notice that the series in Eq. (\ref{F-series-result}) represents a Taylor expansion of an exponential.
After normalizing the result for a small $\kappa$, we obtain the transition time distribution in a form
\begin{align}
\label{F-result}
F(t) = (2k G/\kappa)
\exp \left( -kt - (2G/\kappa)e^{-kt} \right)\,.
\end{align}
The form of the obtained expression is the same as the one that would be obtained in a 1D transition
time problem. The propagator for the 1D FP equation in the weak-noise limit has been calculated in
Ref. \onlinecite{Weiss} by using a semiclassical approximation for the corresponding path integral,
yet we are not aware of the simple form (\ref{F-result}) being discussed even in connection with a purely
one-dimensional transition time problem.

The result (\ref{F-result}) is asymptotically correct for $\kappa\to 0$. In realistic systems,
where the details of the transitions can be observed, $\kappa$ may not be too small.
In the next section we present some numerical results for moderately weak noise,
and demonstrate that our analytical theory still provides an adequate picture, at least on the qualitative level.

\section{Numerical results}
\label{sec:numerical}

In this section we compare the estimates, provided by our approximate theory, with the numerical results
for the transition time p.d.f. We restrict our comparison to 1D examples, since the distribution function
in the weak-noise limit has the same form as in the 1D case.
If our multidimensional potential does not possess any additional length or energy scales,
and the potential curvatures in the transverse directions are not atypically small, the numerical results presented
below are characteristic of a general multidimensional case.

To calculate the transition time p.d.f. numerically we use the following scheme \cite{Risken,ZJZ}.
We first solve the time-dependent FP equation supplemented by the following initial and boundary conditions.
The initial condition corresponds to injection of particles close to the left boundary at $x=x_1+a$:
\begin{eqnarray}
\label{init-cond-inject}
&&
\rho(x,0) = \delta(x - x_1 - a)
\end{eqnarray}
with  a small parameter $a$. The boundary conditions are absorbing:
\begin{eqnarray}
\label{bound-cond}
&&
\rho(x_1,t) = \rho(x_2,t) = 0
\,.
\end{eqnarray}
The solution $\rho(x,t)$ determines the probability flux $\omega(t) = -(\kappa/2)\partial_x \rho(x_2,t)$
through the right boundary, which is proportional to the transition time p.d.f.
Finally, one normalizes the flux and applies the $a\to 0$ limit:
\begin{align}
\label{F-limit}
F(t)=\lim\limits_{a\to 0}\frac{\omega(t)}{\int_0^\infty dt\, \omega(t)}
\,.
\end{align}
Note that the flux through the right boundary vanishes
when the starting point approaches the left boundary, $\lim_{a\to 0} \omega(t) = 0$.

\begin{figure}[t]
  \begin{center}
      \includegraphics[width=0.4\textwidth]{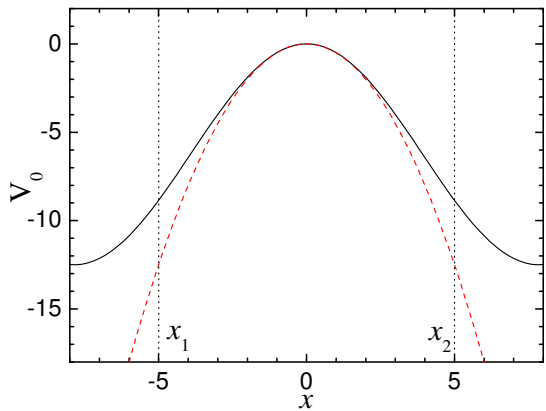}
  \end{center}
  \caption{Two potentials used to illustrate the theory: an inverted parabola $V_0 = x^2/2$
  and an anharmonic potential $V_0=6.25\sin(x/2.5)$. Transition boundaries are $x_1=-5$ and $x_2=5$.}
  \label{fig:potentials}
\end{figure}

\begin{figure}[t]
  \begin{center}
      \includegraphics[width=0.45\textwidth]{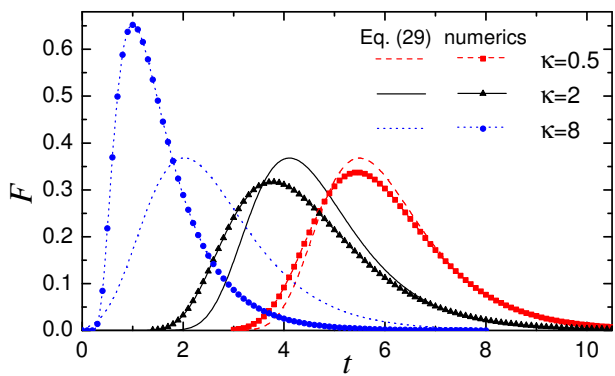}
  \end{center}
  \caption{Transition time distributions for the anharmonic potential shown in
  Fig. \ref{fig:potentials}: from Eq. (\ref{F-result}) (lines) and numerical results
  (lines with symbols).
  Distributions are shifted to longer times as the noise strength decreases.}
  \label{fig:pdfs}
\end{figure}

The comparison of the transition time p.d.f. obtained numerically and from Eq. (\ref{F-result}) is shown
in Fig. \ref{fig:pdfs} for an anharmonic potential sketched in Fig. \ref{fig:potentials}.
We use the corresponding harmonic potential (also shown in Fig. \ref{fig:potentials}) to estimate the source of
discrepancies in the transition time p.f.s.'s.
Since the propagator of the unrestricted FP equation
(which in 1D coincides with the matrix element in Eq. (\ref{F-matrix-element})) is known exactly
for the harmonic potential, it is easy to check
the approximation (\ref{F-matrix-element}) for the transition time p.d.f. by the relevant matrix element,
as well as the subsequent approximation for the the matrix element by the spectral expansion
(\ref{F-series-result}). Both approximations are of the WKB type: they imply weak fluctuations.
In fact, we observe that the accuracies of the two approximations are similar.
The final expression (\ref{F-result}) even turns out to be slightly better than the approximation
(\ref{F-matrix-element})
when it comes to such characteristics of the transition time distribution as the most probable and average
transition times. A noticeable deviation of the expression (\ref{F-result}) from the numerical
results occurs at stronger noise, when the former overestimates the probability of both the shortest and the
longest transition times. At moderately weak noise our expression (\ref{F-result}) correctly
approximates the overall shape of the distribution function, as well as its parameters.

\begin{figure}[t]
  \begin{center}
      \includegraphics[width=0.4\textwidth]{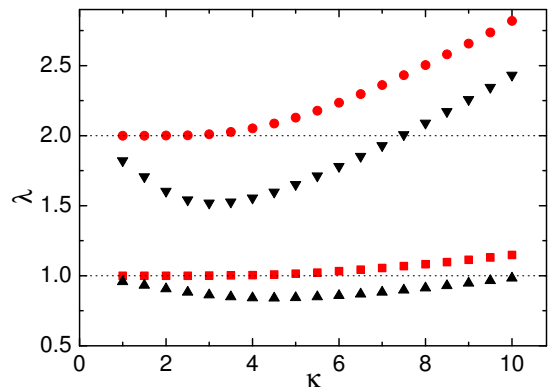}
  \end{center}
  \caption{Two lowest eigenvalues of $-{\cal L}$ with absorbing boundary conditions for
   two potentials shown in Fig. \ref{fig:potentials}, harmonic (red squares and circles)
  and anharmonic (black triangles). }
  \label{fig:eigenvalues}
\end{figure}

Figure \ref{fig:eigenvalues} shows the first two eigenvalues of the FP operator $\cal L$
with the absorbing boundary conditions.
The lowest eigenvalue of $-\cal L$ determines the asymptotic decay rate
of the transition time p.d.f. The second eigenvalue, in particular, estimates the region where the asymptotic
regime is valid. The zero-noise limits for the eigenvalues, which enter
the expansion (\ref{F-series-result}), are shown in the figure with dotted lines.

The spectrum of the FP operator becomes strongly dependent on the boundaries $x_1$ and $x_2$
when they exit the WKB sub-domains surrounding the saddle point. This is reflected in the overall
shape of the transition time p.d.f. For instance, if the starting point $x_1$ moves to the left
beyond the fluctuation region near the potential minimum at $x=y_1$, the corresponding
``transition time'' obviously includes exponentially long (if noise is weak) residence near
the potential minimum; the resulting ``transition time'' distribution resembles the life
time distribution with the exponentially small asymptotic decay rate.

\begin{figure}[h!]
  \begin{center}
      \includegraphics[width=0.4\textwidth]{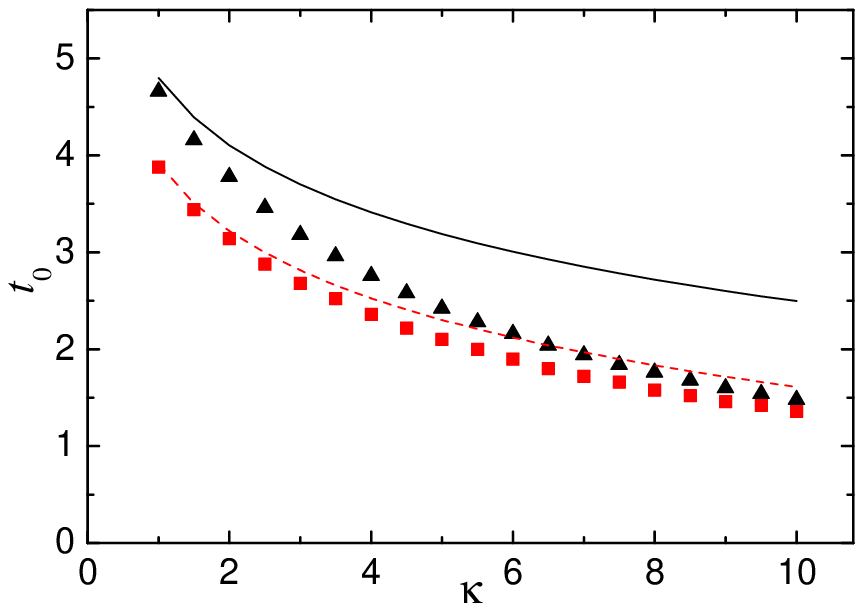}
      \includegraphics[width=0.4\textwidth]{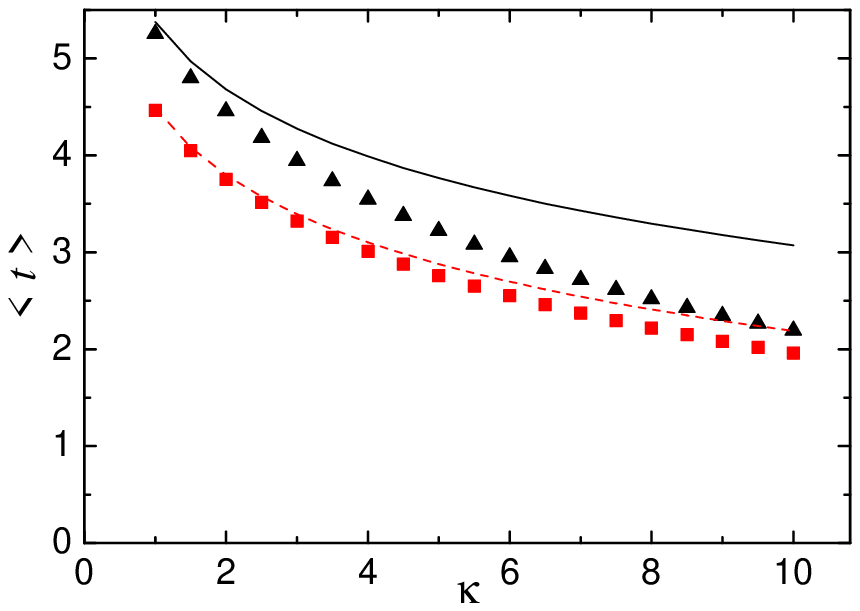}
  \end{center}
  \caption{Comparison of characteristic times obtained from numerical solution
  of the FP equation (lines) and from the approximate p.d.f. (\ref{F-result}) (symbols) for two potentials
  shown in Fig. \ref{fig:potentials}: harmonic (dashed line, red squares)
  and anharmonic (solid line, black triangles). Top: the most probable transition times;
  bottom: the average transition times. }
  \label{fig:times}
\end{figure}

The shifts of the distributions to longer times, which we see in Fig. \ref{fig:pdfs},
can be easier observed when analyzing the behavior of the most probable and average transition times.
These characteristics of the p.d.f. are presented in Fig. \ref{fig:times}.
The agreement between the approximation (\ref{F-result}) and the numerical results
is satisfactory even for moderate noise strengths. The quantitative agreement naturally improves
when the noise becomes weaker compared to the deterministic force outside the saddle point
vicinity.

The result in Eq. (\ref{F-result}), asymptotically correct in the weak-noise limit,
is valid for $t\gg k^{-1}$. Most importantly, the region of validity includes the vicinity of the maximum
of $F(t)$ at $t_0 = k^{-1} \ln(2G/\kappa)$.
The average transition time is given by $\langle t \rangle = k^{-1} (\gamma_E + \ln(2G/\kappa))$
with $\gamma_E \approx 0.577$ being  the Euler constant.
The transition times only weakly depend on the potential far from the saddle point.
As long as $\kappa$ remains small ($\kappa\ll G$), its variation results  only
in the shift of the time that preserves the shape of the p.d.f.
In particular, the variance of the transition time
$\langle t^2 \rangle -\langle t \rangle^2 = \pi^2/(6k^2)$ does not depend on the barrier height.
Our result for the transition time p.d.f. is independent of the transition direction,
which is a manifestation of the time-reversal symmetry, which is in place also for the case of more general,
not necessarily overdamped, stochastic dynamics \cite{BHB}.

\section{Discussion}
\label{sec:discussion}

We have calculated the distribution of the transition time for a particle moving between
two stable points in a multidimensional potential field, when the noise
is low compared to the potential hill. We showed that the stochastic paths that dominate
the transition process are restricted to a thin tube. This reduces the calculation of
the transition time p.d.f. to an almost 1D problem.

Our main result, expressed by in Eq. (\ref{F-result}), which is asymptotically correct in the weak-noise limit,
is valid for the transition times $t\gg k^{-1}$.
We do not consider shorter transition times.
First of all, they require stronger noise intensity, which can invalidate the description of dynamics
by the overdamped Langevin equation with white Gaussian noise. Second, even if the white Gaussian
noise description is valid, the distribution looses the universal form at shorter times 
(which are more difficult to observe), which substantially complicates utilization of this 
theory for the interpretation of the experimental data in terms of the underlying effective potentials.

The divergence of the transition times in the limit $\kappa\to 0$
is a result of the competition between the deterministic and random forces.
In this regard, in the weak-noise limit the transition problem we consider is similar to the of escape
from the saddle point
\cite{Suzuki, CaroliCaroliRoulet, CaroliCaroliRouletGouyet}.
The asymptotic form of the transition time p.d.f. is also similar to that of the escape time
distribution, $\propto e^{-kt}$. In both cases its form is determined by the lowest eigenvalue
of the FP operator corresponding to the eigenstate localized near the saddle point.

The experimental measurements of the transition times can be utilized to obtain more information
on the effective free energy. In the expression for the transition rate
$\gamma = C(\beta)\exp(-\beta E(\beta))$
both the prefactor and the energy barrier typically depend on the thermal energy $\beta^{-1}=k_BT$.
In particular, in the overdamped model the prefactor for the transition from the potential minimum at $y_1$
through the saddle point at $x=0$ is given by
$C = \sqrt{k k_1\det W(y_1)/\det W(0)}/(2\pi)$, where $k_1 = -\partial_x F_0(y_1)$,
$k = \partial_x F_0(0)$, and $W(x)$ characterize the potential $V(x,\bm\zeta)$.
The latter is is obtained from the physical potential by rescaling the coordinates with the mobility tensor;
both the physical potential and the mobility tensor being temperature-dependent.
Therefore, $E(\beta)$ may not be found by just measuring the rates at different temperatures.
In biological molecules, for example,
the temperatures of interest often lie in a narrow interval close to the first-order phase transition,
which leads to a pronounced temperature dependence of  $C(\beta)$ and $E(\beta)$.
In many cases (such as folding of small RNA fragments) a typical lifetime of a metastable state varies
from hundreds milliseconds to several seconds, while the barriers are known to be relatively low
($e^{-\beta E}\sim 10^{-1}\div 10^{-3}$).
This suggests that the mobility
has atypically low values and hence can strongly affect the transition rates.
Thus, the rate measurements alone cannot distinguish the effects of lower mobility in the
preexponential factor $C(\beta)$ from those that originate from higher barriers.
In the simple
overdamped model presented in this paper,
if curvatures of the potential $V(\bm\eta)$ in different directions in the minima and in the saddle
points have similar absolute values, $C$ can be estimated as equal to this typical curvature.
The same quantity determines the asymptotic decay rate in the transition time p.d.f. Thus, a measurement
of the transition time p.d.f. and the transition rate at a single temperature $T=1/(k_B\beta)$ allows
the barrier $E(\beta)$ to be estimated. Additional information on the potential landscape and the
mobility can be obtained from the intra-well relaxation time \cite{NGHS}.

In this manuscript we studied in detail the properties of transitions between two metastable states. The results
for the transition time distribution were obtained in the limit of vanishing noise. In realistic systems
where the transitions can be observed, noise cannot be too weak. We expect that although under such circumstances
our results become less precise, they still adequately describe the general features of the transition time distribution.

The developed approach can be extended to a multidimensional system with multiple metastable states. in the weak noise
(low temperature) limit, the resulting model is reduced to a Markov chain process, defined on a graph,
whose links represent the saddle points (transition states), and are described by the transition time distributions.
An interesting example of such a system would be a hydrogen-bond network \cite{PVoth} whose dynamics can be observed
using 2D infrared spectroscopy  \cite{Hochstrasser, CBHDCNEM, KHKSH, Tokmakoff1, Tokmakoff2,CGJT, PMF}.
Although the H-bond rearrangements in water happen on much faster time scales compared to the conformational
transitions in typical biomolecules,
the corresponding transition paths can be still observed indirectly by studying evolution of
the vibrational frequency, available via 2D spectroscopic measurements \cite{GevaSkinner,JK}.
Application of the model described in this paper to hydrogen-bond networks will help to connect
the transition mechanisms and transition state properties in these complex multidimensional systems
to the observed spectroscopic signatures.

\acknowledgments This material is based upon work supported by the National Science Foundation
under Grant No. CHE-0808910.

\appendix
\section{Reduction of the p.d.f. problem to calculation of the unrestricted
propagator}
\label{appendix:reduction}

In this Appendix we show that if the transition occurs between two WKB sub-domains,
the transition time p.d.f. can be approximated by the normalized matrix element of the
evolution operator $e^{{\cal L} t}$, as suggested in Eq. (\ref{F-matrix-element}).
This property expresses the extremely low probability of the deviations from the most probable unrestricted
trajectory in the low-noise limit. The derivation below is performed for the 1D case, since only
the motion along the reference trajectory is relevant in the weak-noise limit for any number of dimensions.
The 1D counterpart for the matrix element, introduced in the main text, is the propagator $P(x,t)$ for
the unrestricted FP equation (i.e., the equation without specific boundary conditions),
whose spectral expansion contains only those eigenfunctions that correspond to the transition under study.

Consider the solution $\tilde P(x,t)$ of the FP equation with the absorbing boundary conditions
(\ref{bound-cond}) and the initial condition (\ref{init-cond-inject}).
The transition time p.d.f. can be obtained as the limit of the normalized current through
the right boundary generated by $\tilde P(x,t)$ according to Eq. (\ref{F-limit}):
\begin{align}
\label{app:F-limit}
F(t) = -\lim\limits_{a\to 0}
\frac{\sum_\mu \varphi_{0\mu}(x_1 + a)\partial_x\varphi_{0\mu}(x_2) e^{\mu t}}
{\sum_\mu \mu^{-1}\varphi_{0\mu}(x_1 + a)\partial_x\varphi_{0\mu}(x_2)}
\,,
\end{align}
where the functions $\varphi_{0\mu}$ determine the eigenstates of the FP operator
(${\cal L} e^{-V_0(x)/\kappa} \varphi_{0\mu} = \mu e^{-V_0(x)/\kappa} \varphi_{0\mu}$) with absorbing
boundaries, $\varphi_{0\mu}(x_1)=\varphi_{0\mu}(x_2)=0$. 
We accept without rigorous proof that the series converge and the $a$-dependence disappears when
we approximate $\varphi_{0\mu}(x_1 + a)\approx a \partial_x\varphi_{0\mu}(x_1)$.

Thus, omitting the time-independent factors, we can express $P(x,t)$ and $F(t)$ in terms of eigenfunction
expansions:
\begin{align}
\label{app:P}
&
P(x_2,t) \propto
\sum_\lambda \psi_{0\lambda}(x_1)\psi_{0\lambda}(x_2)
e^{\lambda t}
\,,
\\
\label{app:F}
&
F(t) \propto
\sum_\mu \partial_x\varphi_{0\mu}(x_1)\partial_x\varphi_{0\mu}(x_2)
e^{\mu t}
\,.
\end{align}

Now the problem is reduced to finding a relation between the eigenfunctions with and without
absorbing boundaries. First, we recall that the eigenfunctions $\psi_{0\lambda}$ without absorbing
boundaries are localized in the harmonic region near the saddle point and contain
only the decaying WKB ``waves'',
\begin{align}
\label{app:WKB-onewave}
&
\psi_{0\lambda_n} = A_ne^{-\phi(x)}/\sqrt{p(x)}
\\
&
\phi(x) = \int^x_{x_0} dz p(z)\,,
\\
\label{app:WKB-onewave-1}
&
p(x)=\kappa^{-1} \sqrt{(\partial_x V_0)^2 - \kappa\partial^2_x V_0
+ 2\kappa\lambda}\,.
\end{align}

The functions $\varphi_{0\mu}$ are also localized near the saddle point. They do need to
contain the growing WKB ``waves'' that cancel out the decaying ``waves'' at the boundaries
$x_1$ and $x_2$.
However, these additional
``waves'' decay rapidly inside the interval $(x_1,x_2)$ and virtually vanish in the saddle-point
harmonic region. Therefore, the eigenvalues of $\varphi_{0\mu}$ are determined in the harmonic region  
and are the same as of the eigenfunctions $\psi_{0\lambda_n}$: $\mu = -nk$. 
Moreover, since normalization in the harmonic region determines the amplitudes of the
decaying ``waves'' in the WKB domain, they are equal 
in the corresponding $\varphi_{0\lambda_n}$ and $\psi_{0\lambda_n}$.
Thus, in the WKB domain we obtain the following eigenfunctions with absorbing boundaries:
\begin{align}
\label{app:WKB-twowaves}
&
\varphi_{0\lambda_n}(x) = A_ne^{-\phi(x)}/\sqrt{p(x)} + B_ne^{\phi(x)}/\sqrt{p(x)}
\,,
\end{align}
where $A_n$ and $p(x)$ are the same coefficient as in the corresponding $\psi_{0\lambda_n}$ 
in Eq. (\ref{app:WKB-onewave}).
The coefficients $B_n$ are generally different in the two WKB sub-domains. At $x>0$
the amplitude $B_n$ of the growing ``wave'' is found from the boundary condition 
$\varphi_{0\lambda_n}(x_2)=0$:
\begin{align}
\label{app:bound-cond-waves}
&
B_n e^{\phi(x_2)} = -A_n e^{-\phi(x_2)}
\,.
\end{align}
From Eq. (\ref{app:WKB-twowaves}) in the WKB domain we obtain
\begin{align}
\label{app:dxWKB-twowaves}
&
\partial_x\varphi_{0\lambda_n}(x) = -\sqrt{p(x)}(A_ne^{-\phi(x)} - B_ne^{\phi(x)})
\,.
\end{align}
Finally, combining Eqs. (\ref{app:WKB-onewave}), (\ref{app:bound-cond-waves}) and (\ref{app:dxWKB-twowaves}),
we find that the WKB eigenfunctions with and without absorbing boundaries are related by
\begin{align}
\label{app:dxWKB-relation}
&
\partial_x\varphi_{0\lambda_n}(x_2) = -2 p(x_2) \psi_{0\lambda_n}(x_2)
\,.
\end{align}
A similar relation can be found at $x_1$.
The dependence of $p(x)$ on $\lambda$ can be neglected in the weak-noise limit, since the series
over the eigenvalues rapidly converges for relevant transition
times. The easiest way to verify that is to analyze the final expression (\ref{F-result}).
Therefore, comparing Eqs. (\ref{app:P}) and (\ref{app:F}), we obtain the following weak-noise relation
\begin{align}
\label{app:FP}
&
F(t) \propto P(x_2,t)
\,.
\end{align}


\begin{thebibliography}{99}

\bibitem{Gardiner} C. Gardiner, \textit{Handbook of Stochastic Methods for Physics,
Chemistry and the Natural Sciences} (Springer, 2nd edition, 1996).

\bibitem{HTB} P. H\"anggi, P. Talkner, and M. Borkovec, Rev. Mod. Phys. {\bf 62}, 251 (1990).

\bibitem{MO} W. E. Moerner and M. Orrit, Science {\bf 283}, 1670 (1999).

\bibitem{BJS} E. Barkai, Y. J. Jung, and R. Silbey, Ann. Rev. Phys. Chem. {\bf 55}, 457 (2004).

\bibitem{Ritort} F. Ritort, J. Phys.: Condens. Matter {\bf 18}, R531 (2006).

\bibitem{93EvansCohenMorris} D. J. Evans, E. G. D. Cohen, and G. P. Morris, Phys.
Rev. Lett. {\bf 71}, 2401 (1993).

\bibitem{95GallavottiCohen} G. Gallavotti and E. D. G. Cohen,
J. Stat. Phys. {\bf 80}, 931 (1995).

\bibitem{97Jarz} C. Jarzynski, Phys. Rev. Lett. {\bf 78}, 2690 (1997).

\bibitem{00Crooks} G. Crooks, Phys. Rev. E {\bf 61}, 2361 (2000).

\bibitem{05Seifert} U. Seifert, Phys. Rev. Lett. {\bf 95}, 040602 (2005).

\bibitem{06CCJarz} V. Chernyak, M. Chertkov, and C. Jarzynski,
J. Stat. Mech. P08001 (2006).

\bibitem{05CCJarz} V. Chernyak, M. Chertkov, and C. Jarzynski, Phys. Rev.
E {\bf 71}, 025102 (2005).

\bibitem{LS99} J. L. Lebowitz and H. Spohn, J. Stat. Phys. {\bf 95}, 333 (1999).

\bibitem{98Kurchan} J. Kurchan, J. Phys. A: Math. Gen. {\bf 31}, 3719 (1998).

\bibitem{HSch07} R. J. Harris and G. M. Schutz,
J. Stat. Mech.: Theory Exp. P07020 (2007).

\bibitem{TurCCP07} K. Turitsyn, M. Chertkov, V. Y. Chernyak, and A. Puliato,
Phys. Rev. Lett. {\bf 98}, 180603 (2007).

\bibitem{07CKLTur} M. Chertkov, I. Kolokolov, V. Lebedev, and K. Turitsyn,
J. Fluid. Mech. {\bf 531}, 251 (2005).





\bibitem{AC05} J. S. Aldridge and A. N. Cleland, Phys. Rev. Lett. {\bf 94}, 156403 (2005).

\bibitem{LMMMS} D. G. Luchinsky, R. S. Maier, R. Mannella, P. V. E. McClintock, and D. L. Stein,
Phys. Rev. Lett. {\bf 79}, 3109 (1997).

\bibitem{LMcCD} D. G. Luchinsky, P. V. E. McClintock, and M. I. Dykman,
Rep. Prog. Phys. {\bf 61}, 889 (1998).

\bibitem{LLZTHC} T.-H. Lee, L. J. Lapidus, W. Zhao, K. J. Travers, D. Herschlag, and S. Chu,
Biophys J. {\bf 92}, 3275 (2007).

\bibitem{SE} B. Schuler and W. A. Eaton, Curr. Opin. Struct. Biol. {\bf 18}, 16 (2008).


\bibitem{TLRTJDH} D. S. Talaga, W. L. Lau, H. Roder, J. Tang, Y. Jia, W. F. DeGrado,
and R. M. Hochstrasser, Proc. Natl. Acad. Sci. USA {\bf 97}, 13021 (2000).

\bibitem{NGHS} D. Nettels, I. V. Gopich, A. Hoffmann, and B. Schuler,
Proc. Natl. Acad. Sci. USA {\bf 104}, 2655 (2007).

\bibitem{HZRD} J. Hales, A. Zhukov, R. Roy, and M. I. Dykman Phys. Rev. Lett. {\bf 85}, 78 (2000).


\bibitem{08CDS} H. B. Chan, M. I. Dykman, and C. Stambaugh, Phys. Rev. Lett. {\bf 100},
130602 (2008);  Phys. Rev. E {\bf 78}, 051109 (2008).


\bibitem{Tokmakoff2} J. D. Eaves, J. J. Loparo, C. J. Fecko, S. T. Roberts, A. Tokmakoff, and P. L. Geissler,
Proc. Natl. Acad. Sci. USA {\bf 102}, 13019 (2005).

\bibitem{CBHDCNEM}  M. L. Cowan,  B. D. Bruner, N. Huse, J. R. Dwyer, B. Chugh, E. T. J. Nibbering, T. Elsaesser,
and R. J. D. Miller, Nature {\bf 434}, 199 (2005).

\bibitem{Tokmakoff1}  J. J. Loparo, S. T. Roberts,  and  A. Tokmakoff, J. Chem. Phys. {\bf 125} 194522 (2006).

\bibitem{KHKSH} C. Kolano, J. Helbing, M. Kozinski, W. Sander,  and P. Hamm,
Nature {\bf 444}, 469 (2006).

\bibitem{CGJT} H. S. Chung , Z. Ganim, K. C. Jones, and A. Tokmakoff,
Proc. Natl. Acad. Sci. USA {\bf 104}, 14237 (2007).

\bibitem{Hochstrasser} R. M. Hochstrasser, Proc. Natl. Acad. Sci. USA {\bf 104}, 14190 (2007).

\bibitem{PMF} S. Park, D. E. Moilanen, and M. D. Fayer,
J. Phys. Chem. B {\bf 112}, 5279 (2008).



\bibitem{DMSSS} M. I. Dykman, P. V. McClintock, V. N. Smelyanski, N. D. Stein, and N. G. Stocks,
Phys. Rev. Lett. {\bf 68}, 2718 (1992).

\bibitem{MaierStein} R. S. Maier and D. L. Stein, SIAM J. Appl. Math. {\bf 57},  752 (1997).

\bibitem{LN} D. K. Lubensky and D. R. Nelson, Biophys. J. {\bf 77}, 1824 (1999).

\bibitem{ZW} D. M. Zuckerman and T. B. Woolf, J. Chem. Phys. {\bf 116}, 2586 (2002).

\bibitem{MT} A. J. McKane and M. B. Tarlie, Phys. Rev. E {\bf 69}, 041106 (2004).

\bibitem{BHB} A. M. Berezhkovskii, G. Hummer, and S. M. Bezrukov, Phys. Rev. Lett. {\bf 97}, 020601 (2006).

\bibitem{ZJZ} B. W. Zhang, D. Jasnow, and D. M. Zuckerman, J. Chem. Phys. {\bf 126}, 074504 (2007).



\bibitem{SOW} N. D. Socci, J. N. Onuchic, and P. G. Wolynes, J. Chem. Phys. {\bf 104}, 5860 (1996).

\bibitem{KT} D. K. Klimov and D. Thirumalai, Phys. Rev. Lett. {\bf 79}, 317 (1997).

\bibitem{BH} R. B. Best and G. Hummer, Phys. Rev. Lett. {\bf 96}, 228104 (2006).



\bibitem{CaroliCaroliRouletGouyet} B. Caroli, C. Caroli, B. Roulet, and J. F. Gouyet,
J. Stat. Phys. {\bf 22}, 515 (1980).

\bibitem{04TNK}
S. {Tanase-Nicola} and J. Kurchan,
J. Stat. Phys. \textbf{116}, 1201 (2004).

\bibitem{CCMT09} V. Y. Chernyak, M. Chertkov, S. V. Malinin, and R. Teodorescu,
accepted in J. Stat. Phys. (2009), arXiv:0907.3481v2 [cond-mat.stat-mech].

\bibitem{OM53} L. Onsager and S. Machlup, Phys. Rev. {\bf 91}, 1505 (1953).

\bibitem{HuntRoss} K. L. C. Hunt and J. Ross, J. Chem. Phys. {\bf 75}, 976 (1981).


\bibitem{Feynman} R. P. Feynman and A. R. Hibbs, \textit{Quantum Mechanics and Path Integrals}
(McGraw–Hill, New York, 1965).

\bibitem{CaroliCaroliRoulet_pathint} B. Caroli, C. Caroli, and B. Roulet,
J. Stat. Phys. {\bf 26}, 83 (1981).

\bibitem{Weiss} U. Weiss, Phys. Rev. A {\bf 25}, 2444 (1982).

\bibitem{Risken} H. Risken, \textit{The Fokker-Planck Equation} (Springer, 1989).

\bibitem{Graham} R. Graham, Z. Phys. B {26}, 281 (1977);
397 (1977).


\bibitem{LandauerSwanson} R. Landauer and J. A. Swanson, Phys. Rev. {\bf 121}, 1668 (1961).

\bibitem{CaroliCaroliRoulet} B. Caroli, C. Caroli, and B. Roulet,
J. Stat. Phys. {\bf 21}, 415 (1979).

\bibitem{MaierStein2}
R. S. Maier and, D. L. Stein,
Phys. Rev. E \textbf{48}, 931 (1993).

\bibitem{DMRH}
M. I. Dykman, E. Mori, J. Ross, and P. M. Hunt,
J. Chem. Phys. \textbf{100}, 5735 (1994).

\bibitem{DLMcCS}
M. I. Dykman, D. G. Luchinsky, P. V. E. McClintock, and V. N. Smelyanskiy,
J. Chem. Phys. \textbf{77}, 5229 (1996).


\bibitem{Suzuki} M. Suzuki, J. Stat. Phys. {\bf 16}, 477 (1977).
%


\bibitem{PVoth} F. Paesani and G. A. Voth, J. Phys. Chem. B {\bf 113}, 5702 (2009).



\bibitem{GevaSkinner} E. Geva and J.L. Skinner, J. Phys. Chem. B {\bf 101}, 8920 (1997).


\bibitem{JK} T. l. C. Jansen and J. Knoester, J. Chem. Phys. {\bf 127}, 234502 (2007).




\end{thebibliography}
\end{document}